\documentclass{emulateapj}

\slugcomment{Accepted by ApJ}
\shorttitle{A Galactic Jet-Driven Core-Collapse Supernova Remnant}
\shortauthors{LOPEZ ET AL.}

\newcommand{\ltsima}{$\; \buildrel < \over \sim \;$}
\newcommand{\simlt}{\lower.5ex\hbox{\ltsima}}

\newcommand{\ls}{{_<\atop^{\sim}}}

\newcommand{\gs}{{_>\atop^{\sim}}}

\begin{document}

\title{The Galactic Supernova Remnant W49B Likely Originates from a Jet-Driven, Core-Collapse Explosion}

\author{
Laura A. Lopez\altaffilmark{1,2,3}, Enrico Ramirez-Ruiz\altaffilmark{4}, Daniel Castro\altaffilmark{1}, Sarah Pearson\altaffilmark{5} 
}
\altaffiltext{1}{MIT-Kavli Institute for Astrophysics and Space Research, 77 Massachusetts Avenue, 37-664H, Cambridge MA 02139, USA}
\altaffiltext{2}{NASA Einstein Fellow}
\altaffiltext{3}{Pappalardo Fellow in Physics}
\altaffiltext{4}{Department of Astronomy and Astrophysics, University of California Santa Cruz, 1156 High Street, Santa Cruz, CA 95060, USA}
\altaffiltext{5}{DARK Cosmology Centre, Niels Bohr Institute, University of Copenhagen, Juliane Maries vej 30, DK-2100 Copenhagen, Denmark}

\email{lopez@space.mit.edu}

\begin{abstract}

We present results from a 220-ks observation of the galactic supernova remnant (SNR) W49B using the Advanced CCD Imaging Spectrometer (ACIS) on board the {\it Chanrda} X-ray Observatory. We exploit these data to perform detailed spatially-resolved spectroscopic analyses across the SNR with the aim to investigate the thermodynamic properties and explosive origin of W49B. We find substantial variation in the electron temperature and absorbing column toward W49B, and we show that the mean metal abundances are consistent with the predicted yields in models of bipolar/jet-driven core-collapse SNe. Furthermore, we set strict upper limits on the X-ray luminosity of any undetected point sources, and we exclude the presence of a neutron star associated with W49B. We conclude that the morphological, spectral, and environmental characteristics of W49B are indicative of a bipolar Type Ib/Ic SN origin, making it the first of its kind to be discovered in the Milky Way. 

\end{abstract}

\keywords{ISM: Supernova Remnants --- X-rays: ISM --- Supernovae: Individual (W49 B) --- X-rays: ISM}

\section{Introduction}

\begin{figure}
\includegraphics[width=\columnwidth]{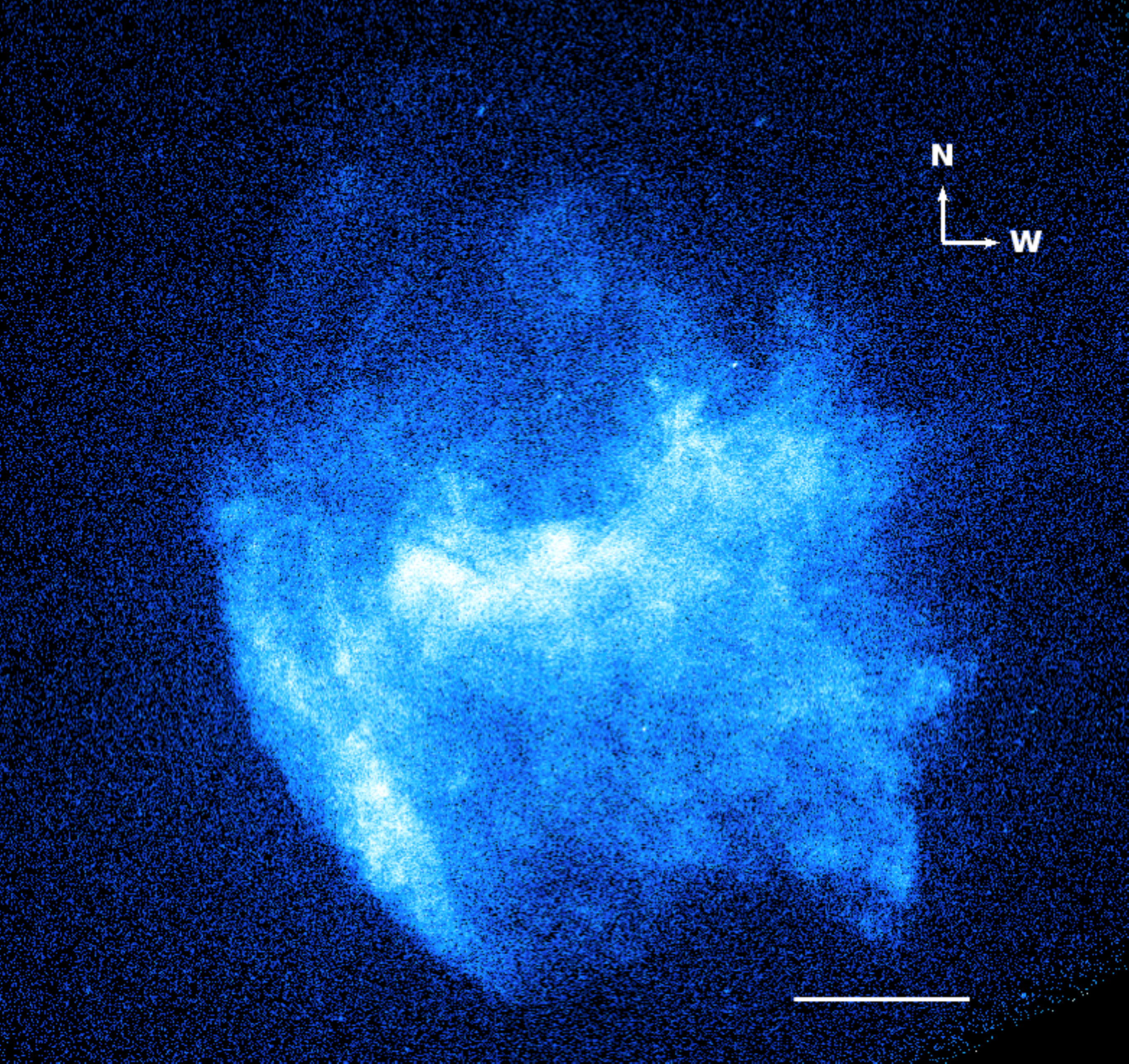}
\caption{Raw X-ray full-band (0.5--8.0 keV) exposure-corrected image of W49B as observed for 220 ks by {\it Chandra} ACIS. Point sources have not been removed to demonstrate the lack of any obvious pulsar or neutron star. Scale bar is 1\arcmin\ in length.}
\label{fig:fullbandimage}
\end{figure}

W49B (G43.3$-$0.2) is one of the most luminous Galactic supernova remnant (SNR) in X-rays (with $L_{\rm X} \sim$ 10$^{38}$ erg s$^{-1}$; \citealt{immler05}) and $\gamma$-rays, as observed by Fermi Large Area Telescope \citep{abdo10}. While radio maps show the SNR has a prominent shell $\sim4\arcmin$ in diameter, the X-ray emission is elongated in a centrally bright, iron rich bar which has two plumes at its eastern and western edges (see Figure~\ref{fig:fullbandimage}; \citealt{hwang00,miceli06,lopez09a}). The X-ray spectrum of W49B has strong emission lines from He-like and H-like ions of several metals with super solar abundances, indicative of their ejecta-dominated  origin.  

Previous studies of W49B have demonstrated the source has several unique morphological and spectral features which are not found in other Galactic SNRs. Foremost, W49B is the only Galactic SNR with large-scale segregation of its nucleosynthetic products \citep{lopez09a,lopez11}: the iron synthesized in the explosion is missing from the Eastern half of the SNR, while the intermediate-mass elements (like silicon and sulfur) are more spatially homogeneous. Secondly, the X-ray spectrum is suggestive of rapid cooling uncommon in young SNRs, such as a strong radiative recombination continuum from iron \citep{ozawa09}. Finally, W49B is one of only two SNRs to date with detected X-ray emission lines from the heavy elements chromium and manganese \citep{hwang00,miceli06}, nucleosynthetic products of explosive silicon burning. 

The physical origin of W49B's unusual traits remains uncertain. Two potential scenarios may account for the barrel shape of W49B \citep{keohane07,miceli08,zhou11}: it could be due to a bipolar/jet-driven explosion of a massive star, or it may be the result of an inhomogeneous ISM around a spherically-symmetric SN. The latter explanation is appealing because it applies to other sources (e.g., G292.0$+$1.8: \citealt{park04}). However, that scenario would require unrealistic ejecta masses in W49B \citep{keohane07}. A bipolar/jet-driven Type Ib/Ic SN origin is consistent with the median abundance ratios from a spatially-resolved spectroscopic analysis using {\it XMM-Newton} observations \citep{lopez09a}, and it may account for the anisotropic distribution of iron since heavy elements are preferentially ejected along the polar axis of the progenitor in these explosions \citep{khok99,mazzali01,err10,couch11}. Based on radio observations, the rate of such events is estimated to be $0.7^{+1.6}_{-0.6}$\% of local Type Ib/Ic SNe \citep{soderberg10}, equating to $\sim$1 per $10^3-10^4$ years per galaxy \citep{lopez11}, with a subset of these explosions being hypernovae or gamma-ray bursts \citep{izz04,pod04}.

In this paper, we investigate the explosive origin and thermodynamic properties of W49B based on a recent 220-ks {\it Chandra} X-ray Observatory Advanced CCD Imaging Spectrometer (ACIS) observation. In Section~\ref{sec:results}, we detail the observations and spatially-resolved spectroscopic analyses of the data. In Section~\ref{sec:discussion}, we discuss the multiwavelength evidence of the explosive origin of W49B and its implications.


\section{Observations and Spectroscopic Analyses} \label{sec:results}

\begin{figure}
\includegraphics[width=\columnwidth]{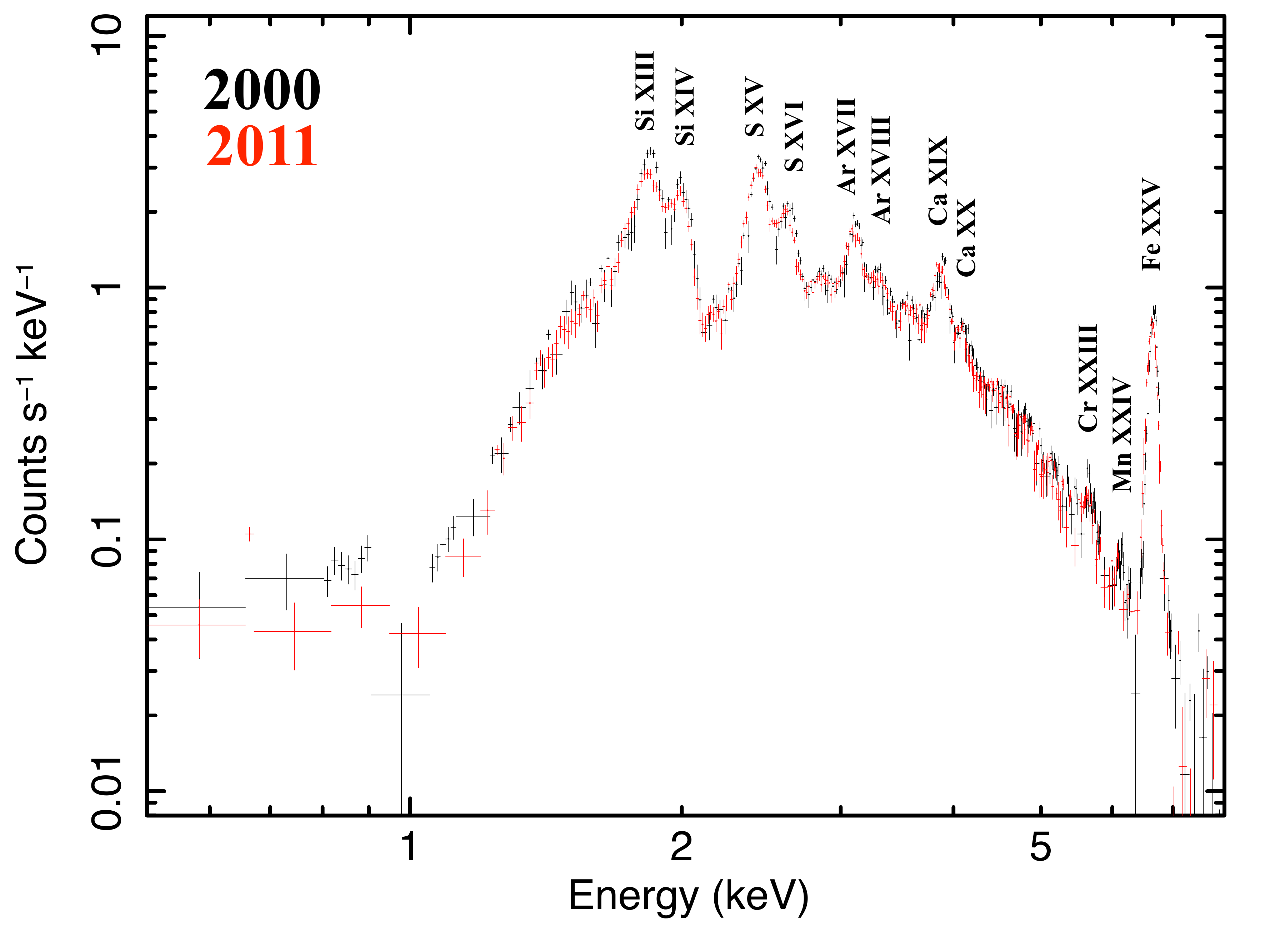}
\caption{Integrated X-ray spectrum from the 220-ks {\it Chandra} observation taken in 2011 (red), with the integrated spectrum from previous {\it Chandra} data taken in 2000. The spectra have numerous prominent emission lines from He-like and H-like ions as well as a strong bremsstrahlung continuum. The high absorbing column toward W49B ($\sim7 \times 10^{22}$ cm$^{-2}$) attenuates the soft X-rays below the Si {\sc xiii} line. In the 11-year interval between observations, the Si {\sc xiii} line flux decreased by $\sim$18\%, and the other emission lines have $\ls$5\% lower fluxes in 2011 compared to their 2000 values.}
\label{fig:globalspectrum}
\end{figure}

W49B was observed for 220 ks with {\it Chandra} ACIS on 18--22 August 2011 with the backside-illuminated S3 chip in the Timed-Exposure Faint Mode (ObsIDs 13440 and 13441). Reprocessed data were downloaded from the {\it Chandra} archive, and analyses were performed using the {\it Chandra} Interactive Analysis of Observations ({\sc ciao}) Version 4.3. We extracted the integrated {\it Chandra} ACIS X-ray spectrum of W49B (Figure~\ref{fig:globalspectrum}) using the {\sc ciao} command {\it specextract} for the central 2.9\arcmin\ around the SNR; a total of $\sim$10$^{6}$ net counts in the 0.5--8.0 keV band were recorded. The spectrum has a strong thermal bremsstrahlung continuum as well as prominent emission lines of silicon, sulfur, argon, calcium, chromium, manganese, and iron.

\begin{figure*}
\includegraphics[width=\textwidth]{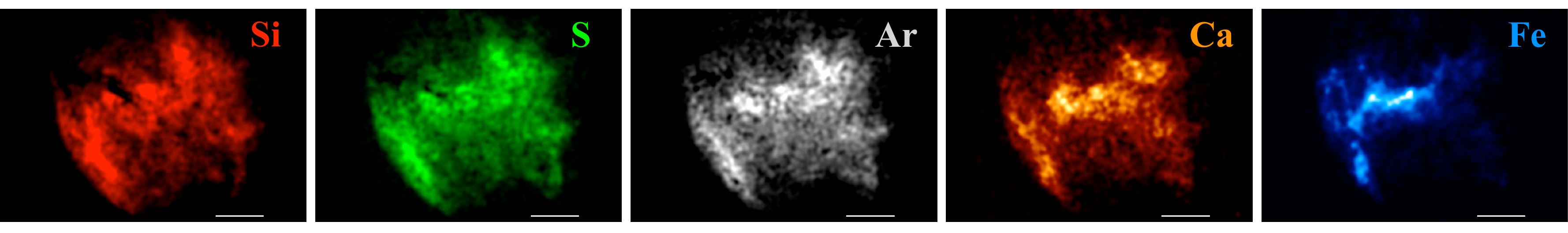}
\caption{Exposure-corrected, continuum-subtracted images of Si {\sc xiii} and Si {\sc xiv}, S {\sc xv} and S {\sc xvi}, Ar {\sc xvii} and Ar {\sc xviii}, Ca {\sc xix} and Ca {\sc xx}, and Fe {\sc xxv}. Images have been smoothed with a Gaussian kernel of $\sigma=$5 pixels. The scale bar is 1\arcmin\ in length; North is up and East is left. Fe {\sc xxv} emission is weak in the West of W49B, while the other ions are more homogeneously distributed. The ``hole'' in the Si image in the Northeast of the SNR is caused by the large absorbing column $N_{\rm H} \sim$10$^{23}$ cm$^{-2}$ there.}
\label{fig:lineimages}
\end{figure*}

We created exposure-corrected images of particular emission features by filtering the data to the narrow energy ranges corresponding to those features: Si {\sc xiii} and Si {\sc xiv} (1.74--2.05 keV), S {\sc xv} and S {\sc xvi} (2.25--2.70 keV), Ar {\sc xvii} and Ar {\sc xviii} (3.0--3.40 keV), Ca {\sc xix} and Ca{\sc xx} (3.72--4.13 keV), and Fe {\sc xxv} (6.4--6.9 keV). We continuum-subtracted these images by creating a continuum image (4.2--5.5 keV), scaling the image to reflect the continuum under the respective lines, smoothing all the images, and then subtracting the scaled continuum images from the line images. The resulting continuum-subtracted images are shown in Figure~\ref{fig:lineimages}. Fe {\sc xxv} is largely absent in the West of W49B, while the other ions are more homogeneously distributed. 

{\it Chandra} had previously observed W49B with ACIS-S3 for 55 ks in 2000. In the 11-year interval between {\it Chandra} observations, the spectra have changed dramatically. In particular, the (unabsorbed) Si {\sc xiii} flux decreased by $\sim$18\%, and the other emission lines have $\ls$5\% lower fluxes in 2011 compared to their 2000 values (see Figure~\ref{fig:globalspectrum}). Based on spatially-resolved spectral analysis, this line flux decrement is the greatest in the West of W49B, whereas the East has the smallest variation. Possible astrophysical explanations for this spectral evolution include SNR expansion or changes in ionization state of the plasma; we defer further analysis and discussion of this spectral evolution to a future paper. 

\subsection{Spectroscopic Analyses}

\begin{figure*}
\includegraphics[width=\textwidth]{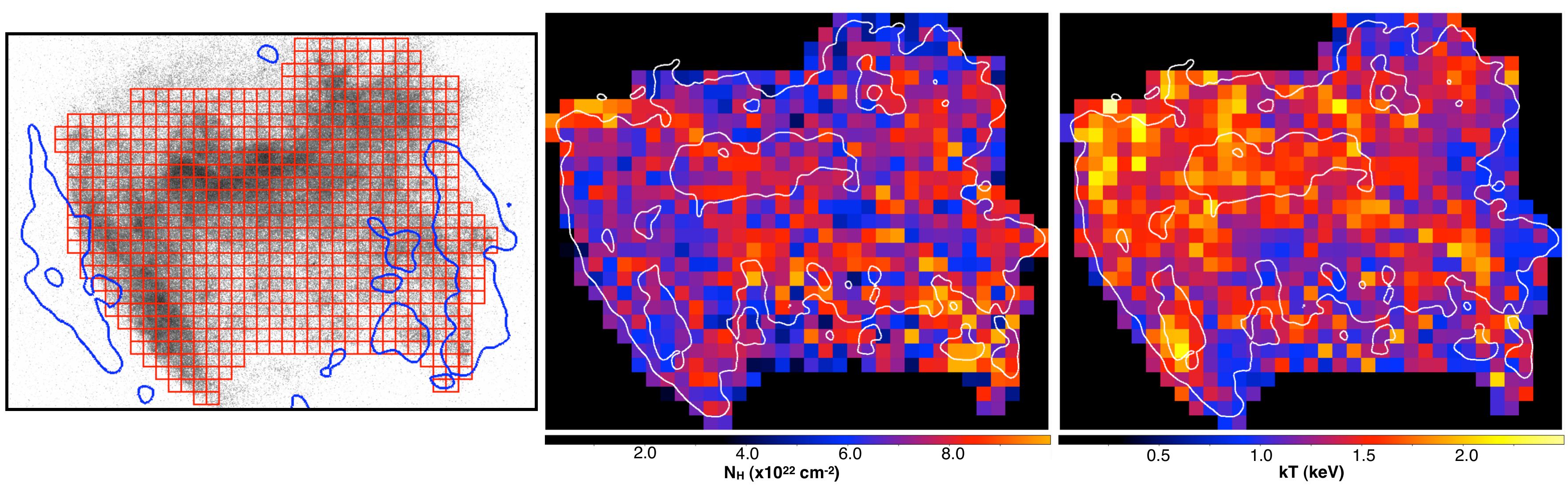}
\caption{Left: Grid of 713, 7.5\arcsec $\times$ 7.5\arcsec\ regions where spectra were extracted. Contours show the location of warm H$_{2}$ (at 2.12 $\mu$m; from \citealt{keohane07}). In the East, the ejecta are impacting the molecular material, while the ejecta are likely freely expanding in the West. Maps of best-fit values from spectral analyses for absorbing column density $N_{\rm H}$ (middle) and electron temperature $kT_{\rm e}$ (right). Overplotted are the X-ray full-band image contours to guide the eye.} 
\label{fig:mapresults}
\end{figure*}

To examine the plasma temperature $kT_{\rm e}$ and column density $N_{\rm H}$ across the SNR, we extracted spectra systematically from a grid of 713 regions of area 7.5\arcsec\ $\times$ 7.5\arcsec\ (see Figure~\ref{fig:mapresults}). Background spectra were extracted from a source-free, 2\arcmin\ diameter region $\sim$1\arcmin\ southeast of the SNR and were subtracted from  the source spectra. Region sizes and locations were determined so that they have $>$400 net counts in the X-ray full band, the minimum necessary to fit the bremsstrahlung shape accurately. Data were grouped to a minimum of 10 counts per bin, and spectra were fit using XSPEC Version 12.7.0 \citep{arnaud96}. We modeled the spectra as an absorbed, variable abundance plasma in collisional ionization equilibrium (CIE) using the XSPEC components {\it phabs} and {\it vmekal} \citep{mewe85,mewe86,liedahl95}. The absorbing column density $N_{\rm H}$, electron temperature $kT_{\rm e}$, abundances (of Si, S, Ar, Ca, and Fe, with solar abundances from \citealt{asplund09}), and normalization were allowed to vary freely during the fitting. Although previous studies employed two CIE plasmas in modeling the X-ray spectra (e.g., \citealt{miceli06,lopez09a}), we find that the use of two components instead of one for our small regions does not statistically significantly improve the reduced chi-squared values of our fits. Therefore, we employ a one-temperature CIE plasma model to fit the spectra of our regions.  

The results of the spatially-resolved spectral analyses of the small, 713 regions are presented in Figure~\ref{fig:mapresults}, where we map the best-fit values of $N_{\rm H}$ and of $kT_{\rm e}$. The fits were statistically good: the mean reduced chi-squared $\chi^{2}$ value was 1.11 with 87 degrees of freedom, with a 1-$\sigma$ range of $\chi^{2}=$1.02--1.19 with 53--101 degrees of freedom. The temperature and absorbing column density vary substantially across the SNR, with values ranging $N_{\rm H} \sim$4--12$ \times10^{22}$ cm$^{-2}$ and $kT_{\rm e} \sim$ 0.7--2.5 keV (with estimated uncertainties of 10\% in both parameters for each region). The $N_{\rm H}$ is enhanced in the northeast and southwest of the SNR, with many regions having $N_{\rm H}$ $\gs8\times$10$^{22}$ cm$^{-2}$. This result is consistent with the work of \cite{lacey01} who showed there may be intervening absorbing material along the line of sight toward W49B in the southwest using low-frequency radio observations. Overall, our measured $N_{\rm H}$ values are greater than those found in previous studies (e.g., \citealt{hwang00} found $N_{\rm H} \approx 5.0\times 10^{22}$ cm$^{-2}$ using integrated {\it ASCA} spectra) because we have employed the revised solar abundances of \citealt{asplund09}. We find a temperature gradient across W49B, with elevated $kT_{\rm e}$ toward the Eastern side of the SNR. The plasma is likely hotter there because it is impacting nearby molecular material to the East (see the contours in Figure~\ref{fig:mapresults}, left; \citealt{keohane07}). 

To assess the abundances of the ejecta metals, we replicated the spatially-resolved spectroscopic analyses described above (i.e., modeled the data as an absorbed, single CIE plasma with variable abundances) using 136 regions of area 15\arcsec $\times$15\arcsec. Regions were selected such that they had $\gs$2500 net full-band counts, the minimum necessary to accurately model the line emission, and the mean number of full-band counts per region was 5500. We set the lower limit on the count rate by producing confidence contours of the abundances for regions of varying count rates, and we found that regions with $\ls$2500 net full-band counts did not yield reliable constraints. We opted to use 15\arcsec $\times$15\arcsec\ regions larger area regions were not adequately fit with a single-temperature plasma (i.e., larger regions may superpose distinct plasmas of different temperatures). The mean reduced chi-squared of the fits was $\chi^{2} =$ 1.50 with 161 degrees of freedom, with a 1-$\sigma$ dispersion range of $\chi^{2} =$1.29--1.71 with 92--232 degrees of freedom. 

\begin{figure*}
\includegraphics[width=\textwidth]{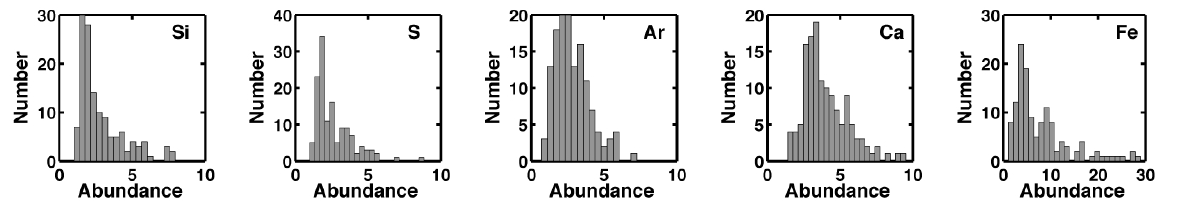}
\caption{Histograms of the abundance ratios by mass obtained from the spatially-resolved spectroscopic analyses across W49B.}
\label{fig:abundhist}
\end{figure*}

In Figure~\ref{fig:abundhist}, we show histograms of the abundances of Si, S, Ar, Ca, and Fe given by our best fits of the 136 regions. The mean abundances of these elements relative to solar by number and their 1-$\sigma$ range are 3.0$^{+1.3}_{-1.5}$, 2.7$\pm$1.1, 2.9$^{+0.9}_{-1.2}$, 4.1$^{+1.5}_{-1.4}$, and 9.1$^{+4.8}_{-5.8}$, respectively. We estimate the uncertainties of the abundances in the fits of individual regions to be $\sim$20--30\%. Therefore, we expect that the large dispersion in abundances reflects the non-uniform distribution of the elements across the SNR. In particular, the large variance in the Fe abundance arises from the fact that the iron is absent in the Western half of W49B, where many regions have Fe at or below their solar values. As the temperatures (from the map in Figure~\ref{fig:mapresults}) in the West are high enough to produce prominent emission from Fe {\sc xxv} if iron is present (see Figure~23 of \citealt{lopez09a}), this result indicates that iron is depleted in the West. 

\section{Discussion of the Explosive Origin of W49B} \label{sec:discussion}

\begin{figure}
\includegraphics[width=0.95\columnwidth]{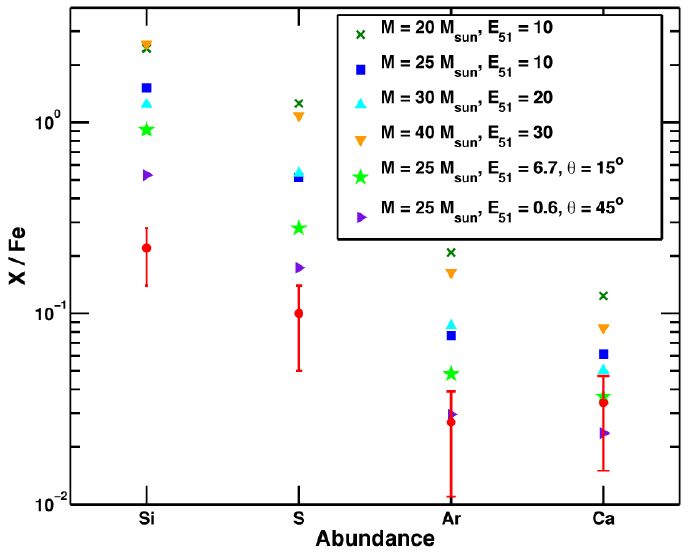}
\caption{Plot of mean abundances (by mass) and their 1$\sigma$ range of values relative to iron of silicon, sulfur, argon, and calcium. For comparison, we show the abundance ratios predicted by models of four spherical explosions (from \citealt{nomoto06}) and two aspherical explosions (from \citealt{maeda03}).}
\label{fig:abundresults}
\end{figure}

In Figure~\ref{fig:abundresults}, we plot the mean abundances by mass and their 1-$\sigma$ range relative to iron of silicon (Si/Fe $\approx$ 0.22$^{+0.06}_{-0.08}$), sulfur (S/Fe $\approx$ 0.10$^{+0.04}_{-0.05}$), argon (Ar/Fe $\approx$ 0.027$^{+0.012}_{-0.016}$), and calcium (Ca/Fe $\approx$ 0.034$^{+0.013}_{-0.019}$), from the fits of the 136 regions. For comparison, we show the abundance ratios predicted by six core-collapse models: four energetic, spherical explosions (from \citealt{nomoto06}) and two aspherical bipolar explosions (25A and 25B from \citealt{maeda03}). Generally, asymmetries in the explosion alter the chemical yield of bipolar explosions relative to typical CC SNe: while typical CC events eject $\sim0.07-0.15 M_{\sun}$ of $^{56}$Ni, bipolar explosions have much greater nickel yields which increase with asphericity, progenitor mass, and explosion energy \citep{umeda08}. The primary difference between the two aspherical models in Figure~\ref{fig:abundresults} is their jet opening angles, $\theta$: 25A has $\theta=15^{\circ}$ and 25B has $\theta=45^{\circ}$. For a fixed kinetic energy, larger $\theta$ means less mechanical energy is released per unit solid angle, affecting the luminosity of the explosion and the nucleosynthetic products. 

The ratios we obtain for W49B agree better with the predictions of the aspherical models than those of the spherical models. The Si/Fe of the aspherical model with larger opening angle, $\theta = 45^{\circ}$, is within $\sim$5$\sigma$ range of our fit values, while the spherical explosions have Si/Fe $\gs$17$\sigma$ away from the measured ratios. The S/Fe fit values are also within 2$\sigma$ of the prediction for the aspherical model with $\theta = 45^{\circ}$ and within 4.5$\sigma$ range of that for the $\theta=15^{\circ}$ aspherical model, whereas the measured S/Fe is $\gs$10$\sigma$ from the ratios of the spherical explosions. For Ar/Fe, the $\theta = 45^{\circ}$ model gives a value roughly equal to our mean estimated Ar/Fe, and the $\theta = 15^{\circ}$ model is $\sim$1.5$\sigma$ from the obtained value. By comparison, the spherical models are $\gs$5$\sigma$ from the measured Ar/Fe in W49B. Finally, both aspherical models give Ca/Fe ratios within the 1$\sigma$ range of those estimated in W49B, and the spherical models give values $\gs$1$\sigma$ above the measured means. Based on the results for each of these four ratios, we find that a bipolar origin of W49B is most consistent with the abundances found across the SNR. 

To exemplify an SNR that has abundances consistent with spherical model predictions, we measured Si/Fe and S/Fe in several regions of G292.0$+$1.8, an SNR thought to be from a spherically-symmetric CC explosion within a complex CSM \citep{park04}. We extracted spectra from 26 regions identified as ejecta substructures in \cite{lopez11} and modeled these data as an absorbed, variable abundance plane-parallel shocked plasma with constant temperature (with XSPEC model components {\it phabs} and {\it vpshock}). We found mean abundance ratios of Si/Fe $\approx$ 1.54 and S/Fe $\approx$ 0.62. These values are consistent with the $M=25~M_{\sun}$ and $M=30~M_{\sun}$ spherical models of \cite{nomoto06}. 

The recent 220-ks {\it Chandra} ACIS observation has facilitated a detailed study of the ejecta and thermodynamic properties across W49B, and the measured abundance ratios are close to those predicted for a bipolar/jet-driven explosion of a 25 $M_{\sun}$ progenitor. By contrast, these ratios are inconsistent with the values for symmetric CC SNe of different explosion energies and progenitor masses. Although the bipolar SN models do not match the measured abundance ratios precisely, the available models of aspherical CC SNe are limited, and the work presented above demonstrates the need for more simulations exploring the effects of asphericity, progenitor mass, and explosion energy on nucleosynthetic yields. 

In addition to its anomalous abundances, W49B has other observational signatures consistent with a bipolar origin. The X-ray line emission of W49B is the most elliptical and elongated when compared to other young, X-ray bright SNRs \citep{lopez09b}, and the barrel shape of W49B is what would be expected from a bipolar SNR if the jet was oriented near the plane of the sky. Additionally, the Fe {\sc xxv} is morphologically distinct from the lower-$Z$ ions: the iron is more asymmetric and is segregated from the lighter elements \citep{lopez09a}. Furthermore, as the temperatures in the locations without prominent Fe {\sc xxv} emission are sufficient to produce that feature, the anomalous distribution of Fe {\sc xxv} likely arises from anisotropic ejection of iron \citep{lopez09a}. The morphological differences of iron from lower-$Z$ elements is consistent with a bipolar origin: the enhanced kinetic energy at the polar axis causes nickel to be more efficiently synthesized there, while perpendicular to the rotation axis, mechanical energy decreases, causing lower-$Z$ elements to also  be ejected more isotropically (e.g., \citealt{mazzali05}).

The environment of W49B is also suggestive of a massive progenitor star that would be necessary for a Type Ic bipolar explosion. \cite{keohane07} imaged W49B at near-infrared wavelengths and found a barrel-shaped structure in the 1.64 $\mu$m [Fe {\sc ii}] which they interpret as coaxial rings from wind-blown material of a massive star. These authors also found warm, shocked H$_{2}$ at 2.12$\mu$m enclosing the SNR to the East and Southwest, while cooler $^{13}$CO gas has been detected at the North and West of W49B \citep{simon01}. This molecular gas is likely from the cloud where the massive progenitor of W49B formed. Moreover, W49B is one of the most luminous SNR in GeV gamma rays \citep{abdo10}, and the brightest SNRs at these energies are interacting with molecular clouds because pion decay of relativistic ions is most efficient when they are accelerated in shocks advancing through dense material. 

\begin{figure}
\includegraphics[width=\columnwidth]{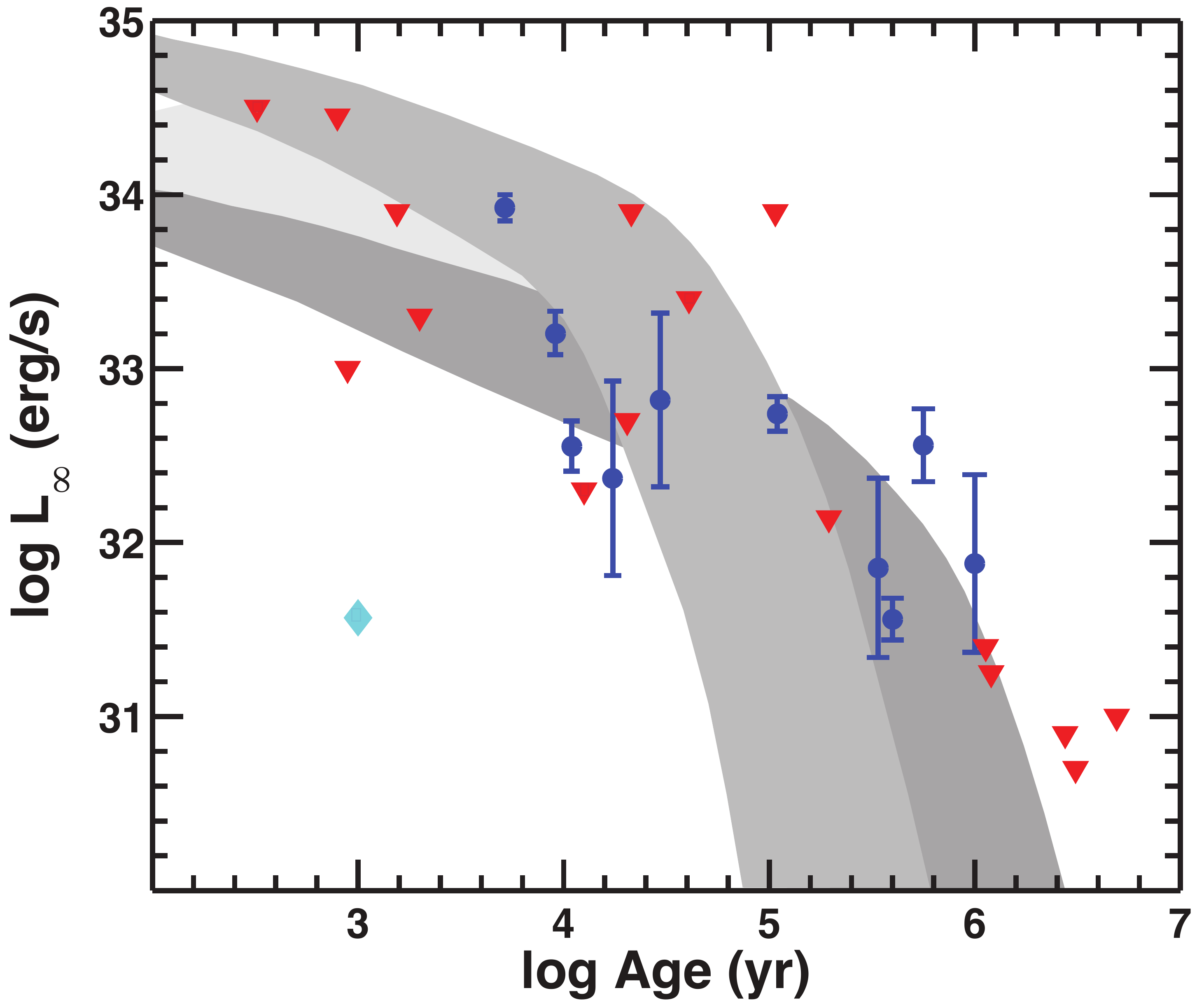}
\caption{Plot of total photon luminosity at infinity $L_{\infty}$ versus age for nearby neutron stars and predictions from models of neutron star cooling. The light blue diamond is our upper limit for a neutron star in W49B; dark blue points are detections and red triangles are upper limits, from \cite{tsu09}. Shaded curves are model predictions of \cite{page06} for neutron star cooling with a heavy element envelope (dark gray), a maximum amount of light elements (medium gray), and an intermediate amount of light elements (light gray). Our upper limit for W49B (${\rm log} L_{\infty} \ls 31.6$) is two to three orders of magnitude below the predictions for a 1000 year old source.}
\label{fig:neutronstars}
\end{figure}

Finally, as the remnant of a bipolar/jet-driven explosion, W49B might be expected to have formed a black hole or a magnetar, and thus no neutron star or pulsar should be detected in the source. From our 220-ks {\it Chandra} observation, we estimate any undetected neutron stars would have an upper limit X-ray luminosity of $L_{\rm X}<3.7 \times 10^{31}$ erg s$^{-1}$ in the 0.1--10 keV band (assuming a 0.5 keV blackbody). Figure~\ref{fig:neutronstars} compares our upper limit in W49B to current detections and upper limits of thermal emission in nearby neutron stars \citep{kaplan04,tsu09}. Furthermore, we plot model predictions of neutron star cooling (from \citealt{page06}). For a 1000 year old point source, the Page models predict total photon luminosities at infinity of ${\rm log} L_{\infty} \sim 33.3-35$. Thus, if W49B hosts a neutron star, our limit requires the central source to be less luminous than those currently known and 2--4 orders of magnitude lower than predictions of neutron star cooling models. 

Furthermore, \cite{gorham96} searched for pulsations in W49B using Arecibo, and they found no pulsations with periods $\gs$0.5 ms down to flux densities of 1.2 mJy at 430 MHz across the entire SNR. Finally, the spectrum of the bright gamma-ray emission detected with Fermi-LAT from W49B is inconsistent with it arising from a pulsar \citep{abdo10}. Given the strict upper limits on the presence of a neutron star/pulsar, it is reasonable to assume that the compact object produced in the explosion might be a black hole. 

Therefore, the morphological, spectral and environmental evidence supports a bipolar/jet-driven SN origin of W49B, and it is the first SNR of this kind to be identified in the Milky Way. Given the estimated rate of these explosions is $\sim$1 per $10^3-10^4$ years per galaxy \citep{lopez11}, it is feasible that a few of the 274 currently known SNRs in the Galaxy \citep{green09} originate from this kind of SN. 

In the future, this result can be verified two ways. If W49B was a bipolar explosion, it should have copious intermediate-mass elements produced through hydrostatic burning of a massive star (like O and Mg). While the high $N_{\rm H}$ toward W49B precludes detection of emission lines from these metals in the optical and X-ray, we predict emission features from elements should be detectable at near-infrared wavelengths (as observed in e.g., Cassiopeia~A: \citealt{gerardy01}). Secondly, if W49B was jet driven, the metals in the central bar should be moving at velocities greater than the more homogeneously distributed, lower-$Z$ plasma. Therefore, using spatially-resolved, high-resolution X-ray spectrometers like Astro-H \citep{astroh}, one could measure the Doppler shifts of the jet plasma and verify its higher speeds than other ejecta material. 

\acknowledgements

We thank Dr. Patrick Slane and Dr. Hiroya Yamaguchi for helpful discussions. Support for this work was provided by National Aeronautics and Space Administration through Chandra Award Number GO2--13003A and through Smithsonian Astrophysical Observatory contract SV3--73016 to MIT issued by the Chandra X-ray Observatory Center, which is operated by the Smithsonian Astrophysical Observatory for and on behalf of NASA under contract NAS8--03060. We acknowledge support from the David and Lucile Packard Foundation and NSF grant AST--0847563. Support for LAL was provided by NASA through the Einstein Fellowship Program, grant PF1--120085, and the MIT Pappalardo Fellowship in Physics. DC acknowledges support for this work provided by NASA through the Smithsonian Astrophysical Observatory contract SV3--73016 to MIT for support of the Chandra X-ray Center, which is operated by the Smithsonian Astrophysical Observatory for and on behalf of NASA under contract NAS8--03060. 


\clearpage

\nocite{*}
\bibliographystyle{apj}
\bibliography{w49b_bib}


\end{document}